\documentclass[12pt]{iopart}
\usepackage{graphicx}
\usepackage{epstopdf}
\usepackage{color}
\usepackage{epsfig}
\usepackage{cite}
\usepackage{amssymb}
\newcommand{\q}{\quad}
\newcommand{\qq}{\qquad}
\newcommand{\bea}{\begin{eqnarray}}
\newcommand{\eea}{\end{eqnarray}}
\newcommand{\ket}[1]{\left|{#1}\right\rangle}

\newcommand{\aver}[1]{\left\langle{#1}\right\rangle}
\newcommand{\expect}[3]{\left\langle{#1}\right|{#2}\left|{#3}\right\rangle}

\newcommand{\inner}[2]{\left\langle{#1}|{#2}\right\rangle}
\newcommand{\trev}{T_{\rm rev}}
\begin{document}
\title{Fractional revivals of superposed coherent states}
\author{M. Rohith$^1$, C. Sudheesh$^1$}
\address{$^{1}$ Department of Physics, Indian Institute of Space Science and Technology, Thiruvananthapuram 695 547, India}
\begin{abstract}
We study the dynamics of superposed wave packets in a specific nonlinear Hamiltonian which models the wave packet propagation in Kerr-like media and the dynamics of Bose-Einstein condensates. We show the dependence of initial wave packet superposition on fractional revival times using analysis based on the expectation values, R\'{e}nyi  entropy and Wigner function. We also show  how the selective identification of fractional revivals  using moments of appropriate  observables depends  on the number of  wave packets present in the initial  state.     
\end{abstract}
\pacs{ 42.50.-p, 03.67.-a}
\maketitle
\section{Introduction}
\paragraph{} Dynamics of the wave packet in a nonlinear media exhibits  revivals and  fractional revivals at specific instants of time, arising from the interference between the stationary states comprising the wave packet.  
The revival phenomena has been investigated both theoretically and experimentally in a wide class of systems \cite{robi}.
An initial well localized quantum state spreads during the propagation and after certain time $T_{rev}$, the revival time, the wave packet localizes again giving rise to quantum wave packet revival. Fractional revival occurs when the initial wave packet evolves into a state that can be described as a collection of mini packets, each of which closely resembles the initial wave packet \cite{aver1}. The fractional revival phenomena has been observed  experimentally in a variety of quantum systems such as Rydberg atomic wave packets \cite{stroud},  molecular vibrational states \cite{marc}, Bose-Einstein condensates \cite{greiner}, etc.
Wave packet isotope separation is closely related to the revivals and fractional revivals, which provides 
the means of suppression of unwanted dispersion of wave packets \cite{aver2}. The collapse and revival
oscillations of first-order coherence were shown to be a sensitive measure of the nearest-neighbor couplings in the extended Bose-Hubbard model \cite{fischer}.

An initial wave packet $\ket{\psi(0)}$ spreads rapidly during the evolution governed by a nonlinear Hamiltonian  and revivals are signaled by the return of the autocorrelation function $A(t)=|\inner{\psi(0)}
{\psi(t)}|^2$ to its initial value of unity.  There are various methods in literature to  identify  and
 analyze fractional revivals. The distinctive signatures of the different fractional revivals of a
suitably prepared initial wave packet are displayed in the mean values and higher moments of appropriate observables \cite{sudh1}.  
Entropy associated with phase  distribution \cite{vaccaro}  or  R\'{e}nyi entropy \cite{romera1} (sum of entropies associated with  position and momentum space probability distributions) is also can be used  to study the formation of macroscopic quantum superposition states.  Wigner function plots  can be used to visualize the revivals and  fractional revivals in phase space.

The universal scenario of revivals described in \cite{aver1} applies to an arbitrary initial superposition states,  including a superposition of several wave packets as well. However, the generic analytical expressions of
\cite{aver1} and the revival phenomena discussed in a wide class of  systems \cite{robi}   are mainly dealing with the arbitrary initial superposition states. To date, revival phenomena of {\it superposed initial wave packets} received less attention in the literature.  Thus, a problem of considerable  interest is to study in detail the revivals and fractional revivals of initial  superposed wave packets. 
For this purpose  we use the example of a specific nonlinear Hamiltonian that is physically relevant in 
at least two important contexts: wave packet propagation in Kerr-like media \cite{milburn,kita}, and the dynamics of BECs \cite{greiner}.  Generation of  discrete superposition of coherent states at fractional revival times in the process of wave packet propagation in Kerr-like media is discussed in \cite{tara,tanas,yurke}.

In most of the earlier studies    \cite{greiner,milburn,kita,tara,tanas,yurke},  the initial state considered is   an   initial  initial coherent state  $\ket{\psi(0)}=\ket{\alpha}$,  where  $\alpha=|\alpha|e^{i\theta}$ is a complex number.  The coherent state $\ket{\alpha}$    is defined as the eigenstate of the annihilation operator 
$a$,  and its  Fock state representation is
\begin{equation}
\ket{\alpha}=e^{-|\alpha|^2/2}\,\sum_{n=0}^{\infty}\,\frac{\alpha^{n}}{\sqrt{n!}}\ket{n}.
\label{cs}
\end{equation}
For ready reference, in section 2  we review some of the relevant results pertained to revival and fractional revival of initial coherent state in the Kerr-like medium. In section 2  we also discuss  how the  fractional revivals are identified  using   expectation values \cite{sudh1}, Wigner function and  R\'{e}nyi entropy \cite{romera1}.

Though  the results obtained in \cite{sudh1,tara,tanas,yurke}  are for   an initial coherent state, they are  applicable to   any initial wave packet  of the form 
\begin{eqnarray}
\ket{\psi(0)}=\sum_{n=0}^{\infty} C_n \ket{n}.
\label{sumn}
\end{eqnarray}
Another example for an initial state of the form given in equation (\ref{sumn}) is m-photon-added coherent state \cite{agar}.  Time evolution of initial m-photon-added coherent state in Kerr-like media shows revival and fraction revivals    at same instants as in the case of  initial coherent state \cite{sudh2}.

This paper will discuss  the fractional revivals    of states  where only every second, third, fourth, etc., expansion coefficient $C_n$ differs from zero, and these states   have not been looked at in detail in  \cite{sudh1,tara,tanas,yurke,sudh2} before. Such states can be obtained by 
  superposing  $l$  coherent states   
 \begin{eqnarray}
 \ket{\psi_{l,h}}=N_{l,h}\,\sum_{r=0}^{l-1}\,\,e^{-i\,2\pi\,r\,h/ l}\,\ket{\alpha\,e^{i\,2\pi\,r/l}},
 \label{generalsuper}
 \end{eqnarray}
where $ {h=0, 1, 2, \dots,l-1}$ and $N_{l,h}$ is a appropriate normalization constant.
The number state representation of the state $\ket{\psi_{l,h}}$ is
 \begin{eqnarray}
\ket{\psi_{l,h}}=l\,N_{l,h}\,\,e^{-|\alpha|^2/2}\,\sum_{n=0}^{\infty} \frac{\alpha^{ln+h}}{\sqrt{(l\,n+h)!}}\,\ket{l\,n+h},
 \label{generalsupernumber}
 \end{eqnarray}
which consists in an arithmetic infinite progression having the state $\ket{h}$ as initial term and 
a common difference, equal to $l$, between successive terms.
 The state $\ket{\psi_{l,h}}$ for a given $l$ and $h$ is also an eigenstate of the operator $a^l$ with eigenvalue $\alpha^l$  \cite{napoli}.
If we set $l=1$ and  $h=0$ in the above  equation we retrieve   the initial coherent state given in equation (\ref{cs}).  For $l=2$, we get  two states which correspond to $h=0$ and $h=1$ and they are called even and odd coherent states respectively. In this paper we study in detail   the initial states $\ket{\psi_{l,h}}$ with $h=0$, denoted by 
\begin{eqnarray}
\ket{\psi_{l}}=N_{l}\,\sum_{r=0}^{l-1} \,\ket{\alpha\,e^{i\,2\pi\,r/l}}.
 \label{generaleven}
 \end{eqnarray}
 The state $\ket{\psi_{l}}$ is  termed as {\it even coherent state of order $l$} \cite{napoli} and its number state representation is 
 \begin{eqnarray}
\ket{\psi_{l}}=l\,N_{l}\,\,e^{-|\alpha|^2/2}\,\sum_{n=0}^{\infty} \frac{\alpha^{ln}}{\sqrt{(l\,n)!}}\,\ket{l\,n}.
 \label{generalevennumber}
 \end{eqnarray}

In sections 3 we study the dynamics of  initial state  $\ket{\psi_{2}}$ ($l=2$), which is a  superposition of two coherent states.    We use the methods described in section 2 to study the effect of superposition of two coherent states on the revival and fractional revivals.  In section 4 we  extend our analysis to the superposition of  three coherent states, i.e., $l=3$ case in equation (\ref{generaleven}).   In section 5 we generalize the  results obtained in sections 3 and 4  to an arbitrary choice of initial superposed wave packets, $\ket{\psi_{l}}$,    and discuss the possibility of experimental manifestations  of our results.

\section{Wave packet dynamics  of coherent state in Kerr-like medium}
\paragraph{} The effective Hamiltonian for the propagation of coherent field in a Kerr medium  \cite{milburn,kita} is
\bea
H=\hbar \chi a^{\dag ^2}{a}^2=\hbar \chi N(N-1)
\label{kerrhamiltonian}
\eea
with $N=a^\dag a$, where $a$ and $a^\dag$ are the usual photon annihilation and creation operator respectively and $\chi$ is  a positive constant. The eigenstates of the operator $N$ are the usual Fock basis ${\ket{n}}$. The numerical value of $\chi$ merely sets the time scale.

Consider the evolution of an initial coherent state $\ket{\alpha}$ through the medium. Such an initial state can be shown to revive periodically with revival time 
$T_{\rm rev}=\pi/\chi$.  Let $\alpha=\nu^{1/2}\,\exp(i\theta)$, where $\nu=|\alpha|^2$ is the mean number of photons in the coherent state.  Without loss of generality, we set $\theta=\pi/4$ throughout this paper.
  In between $t=0$ and $t=T_{\rm rev}$, k-sub-packet fractional revivals occur for the initial wave packet 
at time $t=\pi j /k \chi$, where  $j=1,2,\dots,(k-1)$ for a given value of   $k(>1)$  with a condition that $j$ and $k$ are mutually prime integers. Here onwards we use the notation $(r, s)=1$ to denote the two mutually prime integers $r$ and $s$.
At fractional revival time $t/T_{\rm rev}= j /k$, the initial wave packet splits into $k$ sub-packets which resembles the initial wave packet. Thus, at k-sub-packet fractional revival times discrete  superposition of $k$ coherent states are generated \cite{tara,tanas,yurke}.  For example, at $t= T_{\rm rev}/4$
\begin{eqnarray}
\ket{\psi(t=T_{\rm rev}/4)}&=&\frac{1}{\sqrt{8}}\left[(1-i)\,\ket{\alpha\,e^{i\,\pi/4}}\,+\,\sqrt{2}\ket{\alpha\,e^{-i\,\pi/4}}\right.\nonumber\\
&&\left.-\,(1-i)\,\ket{\alpha\,e^{-i\,3\pi/4}}\,+\,\sqrt{2}\ket{\alpha\,e^{i\,3\pi/4}}\right],
\label{super4}
\end{eqnarray}
which is a superposition of four coherent states.

\begin{figure}[h]
\centering
\includegraphics[scale=0.80]{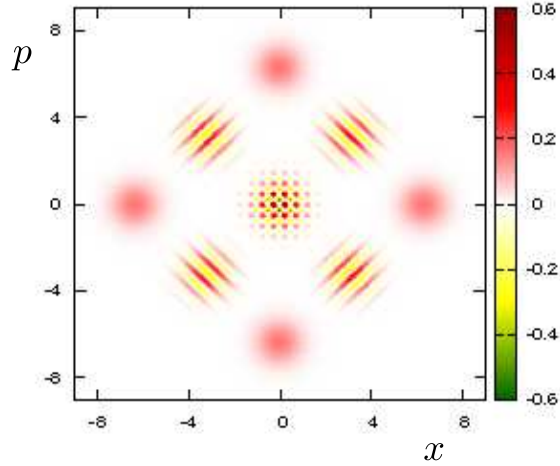} 
\caption{Contour plot of Wigner   function at  $T_{\rm rev}/4$ for an initial coherent state $\ket{\alpha}$ with $\nu=|\alpha|^2=20$. It shows superposition of four coherent states $\ket{\alpha\,e^{i\,\pi/4}},
\ket{\alpha\,e^{-i\,\pi/4}}, \ket{\alpha\,e^{-i\,3\pi/4}}$ and $\ket{\alpha\,e^{i\,3\pi/4}}$ centered at $(0, 2\sqrt{10}), (2\sqrt{10},0)$, $(0, -2\sqrt{10})$, and  $(-2\sqrt{10},0)$, respectively.}
\label{wignercs}
\end{figure}

Consider the operators 
\begin{equation}  
x=\frac{(a+a^{\dagger})}{\sqrt{2}}\quad {\rm and}\quad p=\frac{(a-a^{\dagger})}{i\sqrt{2}}.
\end{equation}
It is convenient to introduce the notation
\begin{eqnarray*}
\alpha=\alpha_1+i\alpha_2=\frac{(x_0+ip_0)}{\sqrt{2}},
\end{eqnarray*}
where $x_0$ and $p_0$ represent the locations of the centers of the  Gaussian wave packets
corresponding to the coherent state $\ket{\alpha}$.  For $\nu=20$, $(x_0, p_0)$ for the states  $\ket{\alpha\,e^{i\,\pi/4}},
\ket{\alpha\,e^{-i\,\pi/4}}, \ket{\alpha\,e^{-i\,3\pi/4}}$ and $\ket{\alpha\,e^{i\,3\pi/4}}$ in equation (\ref{super4})
are $(0, 2\sqrt{10}), (2\sqrt{10},0)$, $(0, -2\sqrt{10})$, and  $(-2\sqrt{10},0)$, respectively. 
Figure \ref{wignercs}
 shows the contour plot of Wigner function at $t=T_{\rm rev}/4$, which clearly shows the superposition
 of four states at the locations mentioned above.

\begin{figure}[h]
\centering
\includegraphics[scale=0.35]{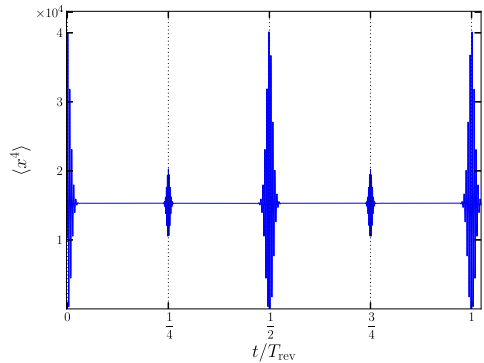} 
\caption{$\aver{x^4(t)}$ as a function of $t/T_{\rm rev}$ for an initial coherent state $\ket{\alpha}$ with $\nu=|\alpha|^2=100$.  In between $t=0$ and $t=T_{\rm rev}$,  $\aver{x^4(t)}$ is constant most of the time except at   fractional revival times  $t=j\,T_{\rm rev}/4$, where $j=1, 2, 3$. At these instants, $4^{\rm th}$ moment of $x$  shows  a rapid variation, which is a signature of 4 and 2-sub-packet fractional revivals.}
\label{x4vst}
\end{figure}

Next, we discuss the manifestations of fractional revivals in moments of observables. 
The time dependence 
of all moments of $x$ and $p$ can be obtained from the general 
result \cite{sudh1}
\begin{eqnarray}
\aver{a^{\dagger r}\,a^{r+s}} &= &\expect{\psi(t)}{a^{\dagger r}\,a^{r+s}}{\psi(t)}= \alpha^{s}\,\nu^{r}
e^{-\nu\,(1-\cos \,2 s \chi t)}\,\\\nonumber
&\times&\exp \left[
-i \chi \big(s(s-1) + 2rs\big)\,t - i\nu \,\sin\,2s \chi t\right],
\label{nthmoment}
\end{eqnarray}
where $r$ and $s$ are non-negative integers.  The time dependence of  $k^{{\rm th}}$  moment of $x$ and $p$  is strongly controlled 
by the factor $\exp[-\nu(1-\cos\,2k\chi t)], k=1, 2, \dots,$ that modulates the oscillatory term. This acts as  a strong damping factor for large values of $\nu$, except when $\cos (2k\chi t)$ is near unity. This happens 
precisely at revivals (when $t=n \pi/\chi$, an integer multiples of $T_{\rm rev}$) and at the fractional 
revival times $t=(n+j/k)T_{\rm rev}$. Thus, by settings $\nu$ at a suitably large value, we ensure that
the moments are essentially static, bursting into rapid variation at specific instants of time before reverting to quiescence. It can be concluded  that $k$-sub-packet fractional revivals are captured in the $k^{\rm th}$ moment of $x$ or $p$ \cite{sudh1} but not in lower momets. In between $t=0$ and $t=\trev$, $k^{\rm th}$ moment of $x$ or $p$ captures the signature of $2, 3, \dots, k$-sub-packet fractional revivals. 
For example, figure \ref{x4vst} shows the variation of $\aver{x^4}$ versus $t$  for an initial coherent state with $\nu=100$.
The dynamics of $4^{\rm th}$ moment  of  $x$ captures the  signatures of four-sub-packet fractional revivals  at $t/T_{\rm rev}=1/4$ and $3/4$ and the two-sub-packet fractional revival time at  $t/T_{\rm rev}=1/2$ in between  $t=0$ and $T_{\rm rev}$.

\begin{figure}
\centering
\includegraphics[width=5.7in,height=3.2in]{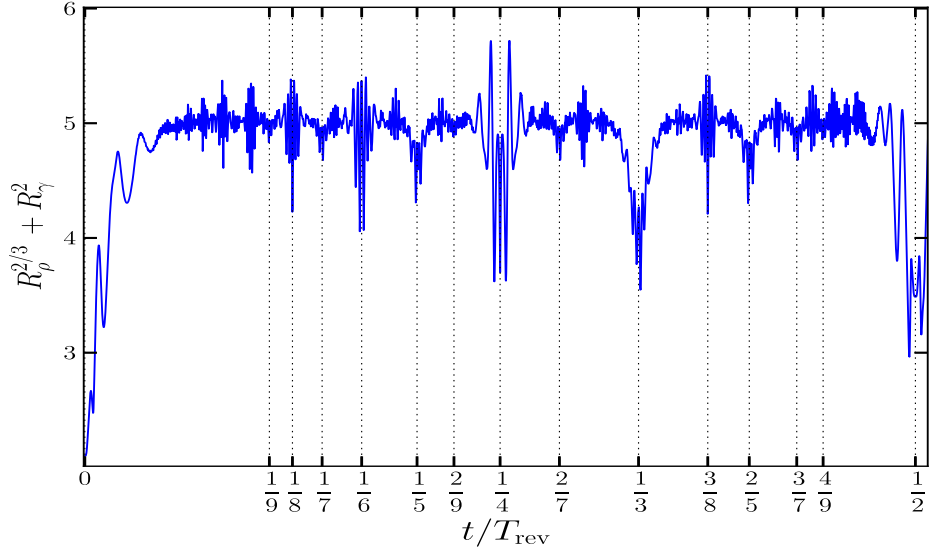} 
\caption{Time evolution of $R_{\rho}^{(2/3)}+R_{\gamma}^{(2)}$ for an initial coherent state with $\nu=|\alpha|^2=35$. The main fractional revivals are indicated by vertical dotted lines.}
\label{renyiCS}
\end{figure}

We also use R\'{e}nyi entropy to analyze the fractional revival phenomena \cite{romera1}. In terms of a generalized probability density $f(x)$ R\'{e}nyi entropy is defined as \cite{birula}
\bea
R_{f}^{(\zeta)} \equiv \frac{1}{1-\zeta} ln \int_{-\infty}^{\infty}[f(x)]^{\zeta} dx \qq \rm for \q 0<\zeta < \infty.
\eea
In terms of probability density in position and momentum spaces,
$\rho(x)=|\psi(x)|^2$ \rm {and} $\gamma(p)=|\phi(p)|^2$, respectively, the R\'{e}nyi uncertainty relation is given by
\bea
R_{\rho}^{(\zeta)}+R_{\gamma}^{(\eta)} \geq -\frac{1}{2(1-\zeta)} ln\frac{\zeta}{\pi}-\frac{1}{2(1-\eta)} ln\frac{\eta}{\pi},
\label{uncertainty}
\eea
with $1/\zeta+1/\eta=2$. As $\zeta\rightarrow 1$ and $\eta\rightarrow 1$ the R\'{e}nyi uncertainty relations reduces to Shannon's, $S_\rho + S_\gamma \geq 1+ln (\pi)$. The entropy function takes local minima at fractional revival times   and thus the signatures of fractional revivals are given by the local minima of $R_{\rho}^{(\zeta)}(t)+R_{\gamma}^{(\eta)}(t)$. Studies based on R\'{e}nyi uncertainty relations for the fractional revivals of infinite square well and  quantum bouncer have been reported  in \cite{romera2, romera3}.
Figure. \ref{renyiCS} displays the time evolution of $R_{\rho}^{(2/3)}+R_{\gamma}^{(2)}$ versus $t/T_{\rm rev}$ for an initial coherent state in Kerr media.  In this figure we have plotted up to $T_{\rm rev}/2$ because it captures all important fractional revivals.  The main fractional revivals are denoted by the vertical dotted lines in figure \ref{renyiCS}.

\section{Evolution of two superposed  coherent states}
\paragraph{} Consider the symmetric   superposition of two  coherent states (set $l=2$ in equation (\ref{generaleven}))
\begin{equation}
\ket{\psi_2}=N_{\rm 2}\left[\ket{\alpha}+\ket{-\alpha}\right],
\end{equation}
where the normalization constant
\begin{equation}
N_{2}=\frac{1}{\sqrt{2}}\left[1+\exp(-2|\alpha|^2)\right]^{-\frac{1}{2}}.
\end{equation}
By setting  $l=2$ in equation (\ref{generalevennumber}) we obtain the  Fock state representation of the even coherent state
 \begin{eqnarray}
\ket{\psi_{2}}=2N_{2}\,\,e^{-|\alpha|^2/2}\,\sum_{n=0}^{\infty} \frac{\alpha^{2n}}{\sqrt{(2\,n)!}}\,\ket{2\,n}.
 \label{general2n}
 \end{eqnarray}

\begin{figure}[!thpb]
\centering
\includegraphics[width=5.8in,height=2.3in]{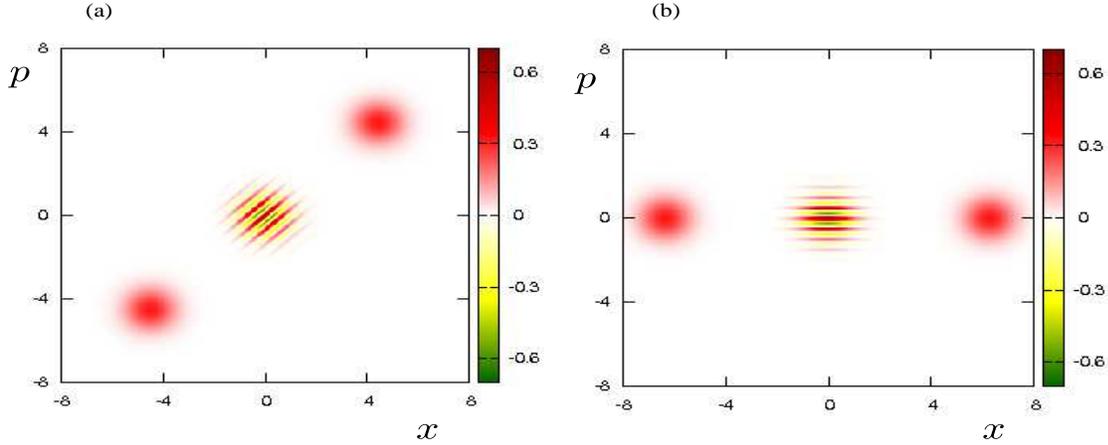}  
\caption{Contour plot of Wigner   function at  (a) $t=0$ and (b) $t=T_{\rm rev}/4$ for an initial even coherent state with $\nu=|\alpha|^2=20$. The unitary time evolution operator at  $t=T_{\rm rev}/4$  rotates the initial even coherent state $45$ degree clockwise direction in phase space.}
\label{Wigner2CST0}
\end{figure}

Consider the dynamics of the initial state $\ket{\psi_2}$ governed by the nonlinear Hamiltonian given in
equation (\ref{kerrhamiltonian}).
In between $t=0$
and $t=T_{\rm rev}$, at $t=jT_{\rm rev}/4$, where $j=1, 2\, {\rm and} \, 3$,  the state is again an even coherent state but rotated in phase space:
\begin{eqnarray*}
\ket{\psi(T_{\rm rev}/4)}&=&N_{2}\,\Big[\ket{\alpha\,e^{-i\pi/4}}+\ket{-\alpha\,e^{-i\pi/4}}\Big]\\
\ket{\psi(T_{\rm rev}/2)}&=&N_{2}\,\Big[\ket{\alpha\,e^{i\pi/2}}+\ket{-\alpha\,e^{i\pi/2}}\Big],\\
\ket{\psi(3T_{\rm rev}/4)}&=&N_{2}\,\Big[\ket{\alpha e^{i\pi/4}}+\ket{-\alpha e^{i\pi/4}}\Big].
\end{eqnarray*}
Figure \ref{Wigner2CST0} shows the plots of the Wigner function for  the states 
$\ket{\psi(0)}=\ket{\psi_2}$ and $\ket{\psi(T_{\rm rev}/4)}$. The unitary time evolution operator at  $t=T_{\rm rev}/4$  rotates the initial even coherent state $45$ degree clockwise direction in phase space.   

\begin{figure}[!htpb]
\centering
\includegraphics[width=2.8in,height=2.7in]{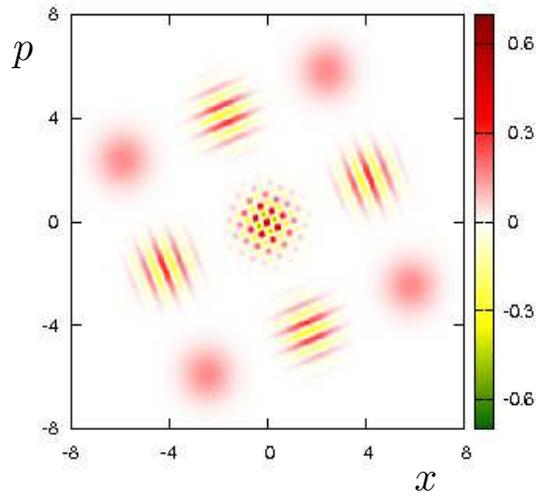} 
\caption{Contour plot of Wigner   function at two-sub-packet fractional revival time $t=T_{\rm rev}/8$ for an initial even coherent state with $\nu=|\alpha|^2=20$.  It shows the superposition of two even coherent states (see equation (\ref{stwoecs})).}
\label{Wigner2CST8}
\end{figure}

Here, $k$-sub-packet fractional revival occurs at time $t=j\,\trev/4k$ where $j=1, 2, \dots, (4k-1)$ for a given value of   $k (>1)$ with $(j, 4k)=1$. At $k$-sub-packet fractional revival time the initial wave packet splits into $k$ sub-packets.  In contrast, we have seen earlier that for an initial coherent state $k$-sub-packet fractional revival occurs  at  $t=j\trev/k$, where   $j=1, 2, \dots, (k-1)$ for a given value of   $k(>1)$  with $(j, k)=1$. For example, two sub-packet fractional revival for an initial even coherent state  occurs at $t=T_{\rm rev}/8$ and 
the state at this time is a superposition of two even coherent state,
\begin{eqnarray}
\ket{\psi(\trev/8)}&=&C_1\,N_{2}\Big[\ket{\alpha\,e^{i\pi/8}}+\ket{-\alpha\,e^{i\pi/8}}\Big]\nonumber\\
&+&C_2\,N_{2}\Big[\ket{\alpha\,e^{-i3\pi/8}}+\ket{-\alpha\,e^{-3i\pi/8}}\Big],
\label{stwoecs}
\end{eqnarray}
where $C_1=(1-i)/2$ and $C_2=(1+i)/2$.
Figure \ref{Wigner2CST8} clearly shows the superposition of two even coherent state at $t=\trev/8$.

\begin{figure}[!htbp]
\centering
\includegraphics[scale=0.4]{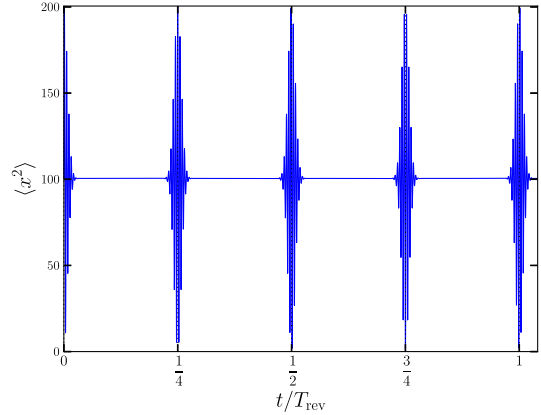} 
\caption{$\aver{x^2(t)}$ as a function of $t/T_{\rm rev}$ for an initial even coherent state $\ket{\psi_2}$ with $\nu=|\alpha|^2=100$. In between $t=0$ and $T_{\rm rev}$,  the second moment of $x$ is constant most of the time except at    $T_{\rm rev}/4, \trev/2$ and  $3\trev/4$. At these instants, $2^{\rm nd}$ moment of $x$  shows  a rapid variation, which is a signature of   wave packet rotation.}
\label{x2vstevencs}
\end{figure}

All odd moments of the operator $x$ and $p$ vanish at all times for the initial even coherent state. The expectation value of $x^2$  at any time can be obtained as explicit functions of $t$ in the form
\begin{eqnarray}
\aver{x^2(t)}&=&2N_2^{2}\,\nu\, \Big[e^{-\nu \left( 1-\cos  4\chi t  \right)}\cos \left( 2\chi t+\nu \sin  4\chi t -\frac{\pi }{4} \right)\nonumber\\
&&+ e^{-\nu \left( 1+\cos \left( 4\chi t \right) \right)}\cos \left( 2\chi t-\nu \sin \left( 4\chi t \right)-\frac{\pi }{4} \right)\Big]+\nu +\frac{1}{2}
\end{eqnarray}
In between $t=0$ and $\trev$, the above expression for $\aver{x^2}$ is static most of the time except at  $t=\trev/4, \trev/2$ and $3\,\trev/4$ for sufficiently large value of $\nu$. Thus, the second moment of $x^2$
captures the signature of wave packet rotation in phase space at $\trev/4, \trev/2$ and $3\,\trev/4$. 
Figure \ref{x2vstevencs} shows the variation of the expectation value $\aver{x^2}$ versus time for the initial even coherent state. 

\begin{figure}[!htbp]
\centering
\includegraphics[scale=0.7]{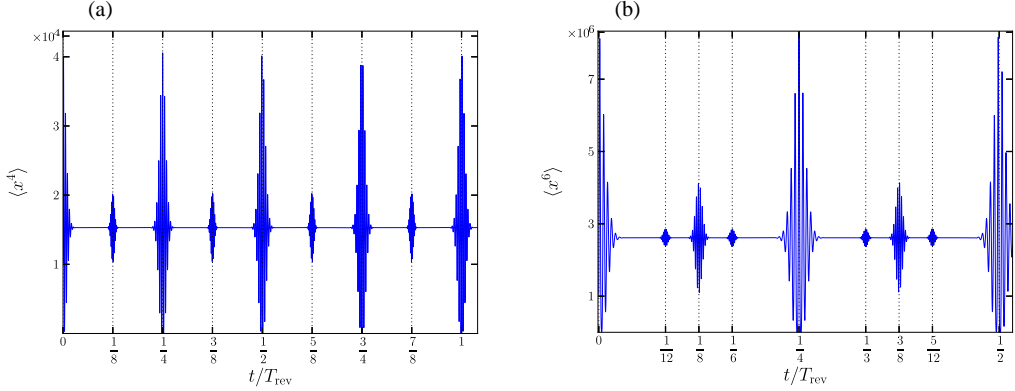} 
\caption{Temporal evolution of higher moments of $x$ for an initial even coherent state $\ket{\psi_2}$  with $\nu=|\alpha|^2=100$. (a) In between $t=0$ and $t=T_{\rm rev}$,  $\aver{x^4(t)}$ is constant most of the time except at   $t=j\,T_{\rm rev}/8$, where $j=1, 2, \dots, 7$.  At these instants, $4^{\rm th}$ moment of $x$  shows  a rapid variation, which is the signature of  two-sub-packet fractional revival  and wave packet rotation.  (b) In this figure we have plotted between $t=0$ and  $\trev/2$ for a  better view. $\aver{x^6(t)}$ is constant most of the time except at   $t=j\,T_{\rm rev}/12$, where $j=1, 2, \dots, 6$.  At these instants, $6^{\rm th}$ moment of $x$  shows  a rapid variation, which is the signature of three and two-sub-packet fractional revivals  and wave packet rotation.}
\label{x2x6forevencs}
\end{figure}

Expressions for the $2k^{\rm th}$ moments of $x$ and $p$ can be deduced readily from the general result
\begin{eqnarray}
\aver{a^{2k}}&=&2N^2_2\, {\alpha}^{2k}\Big\{e^{-\nu(1-\cos 4k\chi t)} \exp\left[ -2ik(2k-1)\chi t-i\nu \sin 4k\chi t\right]  \nonumber\\
&&+ e^{-\nu(1+\cos 4k\chi t)} \exp\left[ -2ik(2k-1)\chi t+i\nu \sin 4k\chi t\right]\Big\},
\end{eqnarray} 
where $k$ is a positive integer. The time dependence of $2k^{\rm th}$ moments of $x$ is
strongly controlled by the factors $\exp[-\nu(1\pm\cos\,4k\chi t)], k=1, 2, \dots,$ that modulates the oscillatory term. In between $t=0$ and $t=T_{\rm rev}$, these factors act as a strong damping factor for large values of $\nu$, except at  fractional revival times $t=j\trev/4k$. It can be concluded that $k$-sub-packet fractional revivals are captured in the $2k^{\rm th}$ moment of $x$ or $p$ but not in lower moments.
These results are illustrated in figures  \ref{x2x6forevencs} (a) and  \ref{x2x6forevencs} (b).
 Figure \ref{x2x6forevencs} (a) shows the temporal evolution of the expectation 
value $\aver{x^4(t)}$. It  shows  rapid oscillations at $t/\trev=j/8$, where $j=1, 2, \dots 7$ in between $t=0$ and $\trev$. Thus,  the fourth moment of $x$ versus time captures the signature of 2-sub-packet fractional revivals at $t/\trev=j/8$ where $j=1, 2, \dots, 7$ with $(j, 8)=1$ and wave packet rotations at $t/\trev=j/4$ where $j=1, 2, 3$. Figure \ref{x2x6forevencs} (b) is a plot of $\aver{x^6(t)}$ versus time which shows the signature of $3$ sub-packet fractional revival.

\begin{figure}[!htpb]
\centering
\includegraphics[width=5.5in,height=2.8in]{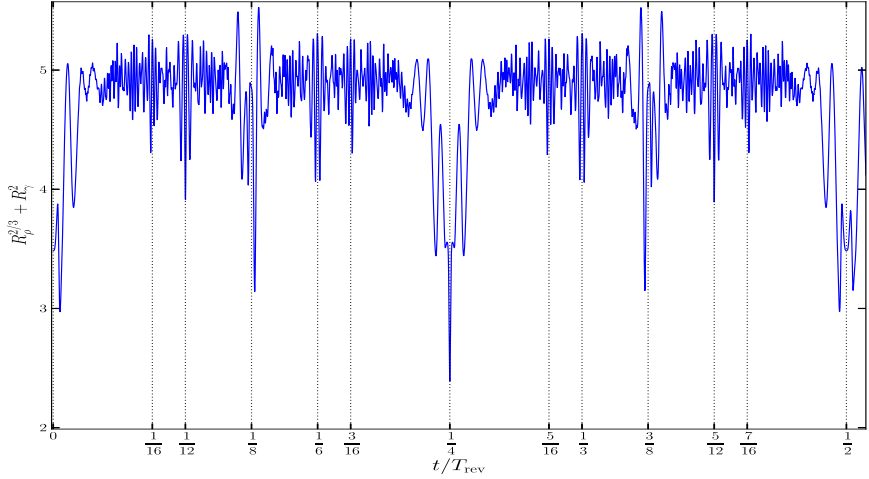} 
\caption{
Time evolution of $R_{\rho}^{(2/3)}+R_{\gamma}^{(2)}$ for an initial even coherent state  with $\nu=30$. The main fractional revivals are indicated by vertical dotted lines.}
\label{entropyevencs}
\end{figure}

We also studied the temporal  evolution of R\'{e}nyi uncertainty relation for the initial even coherent state. Figure \ref{entropyevencs} shows the  R\'{e}nyi uncertainty  versus time for the initial even coherent state with $\nu=30$. Signatures of fractional revivals are indicated by the local minima of  R\'{e}nyi uncertainty.
Our analysis  shows the  clear distinction between  time evolution of initial state of the form $\sum_n\,C_n\ket{n}$ and $\sum_n\,C_n\ket{2n}$. In the next section we study the dynamics of an initial state which is of the form 
$\sum_n\,C_n\ket{3n}$.

\section{Evolution of the initial state of the form $\sum\,C_n\,\ket{3n}$}
\paragraph{}Setting $l=3$ in equation (\ref{generaleven}), we get  superposition of three coherent states
\begin{eqnarray}
\ket{\psi_3}=N_{3}\left[\ket{\alpha}+\ket{\alpha\,e^{i2\pi/3}}+\ket{\alpha\,e^{-i2\pi/3}}\right],
\label{ketpsi3}
\end{eqnarray}
which in Fock space is given by
\begin{eqnarray}
\ket{\psi_3}=3N_{3}\,\,e^{-|\alpha|^2/2}\,\sum_{n=0}^{\infty}\,\frac{\alpha^{3n}}{\sqrt{3n!}}\ket{3n}.
\end{eqnarray}

\begin{figure}[h]
\centering
\includegraphics[width=6.1in,height=2.5in]{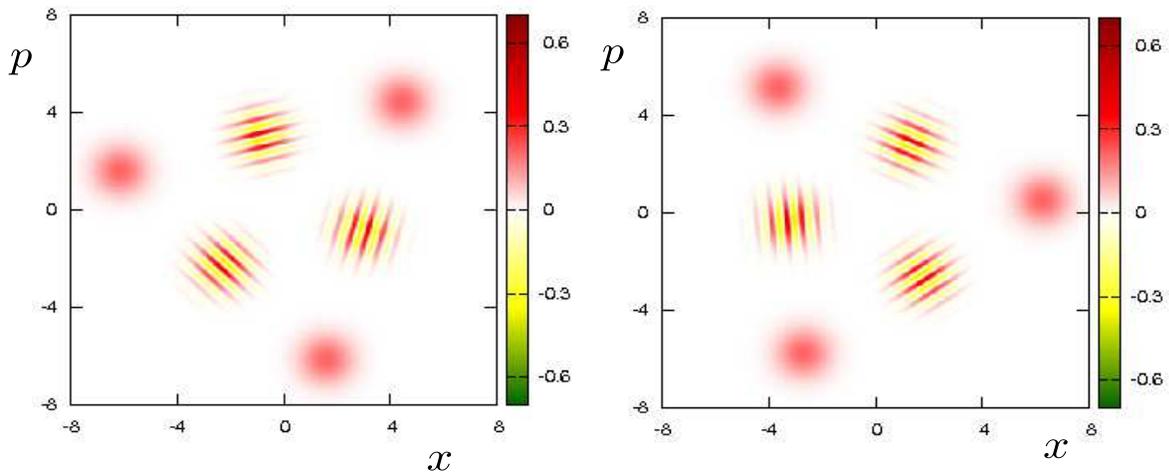} 
\caption{Contour plot of Wigner   function at  $t=0$ (left) and $t=T_{\rm rev}/9$ (right) for the  initial  state $\ket{\psi_3}$  with $\nu=|\alpha|^2=20$.  Both the figures show superposition of three coherent states. The unitary time evolution operator at $t=T_{\rm rev}/9$ rotates the initial state $45$ degree clockwise direction in the phase space.}
\label{Wigner3nt0and9}
\end{figure}

Time evolution of the initial state $\ket{\psi_3}$ shows fractional revivals and rotations at different instants when compared to the initial coherent state and the initial even coherent state. For the initial state 
$\ket{\psi_3}$, the rotations in phase space occur at $t=jT_{\rm rev}/9$, where $j=1, 2, \dots, 8$  in between $t=0$
and $t=T_{\rm rev}$. In the case of initial coherent state there  is no rotation  and for an initial even coherent state rotations occur at $t=jT_{\rm rev}/4$, where $j=1, 2, {\rm and}, 3$.   For example at $t=T_{\rm rev}/9$, the initial state $\ket{\psi_3}$ evolves to
\begin{eqnarray*}
\ket{\psi(T_{\rm rev}/9)}=N_{3}\left[\ket{\alpha\,e^{-i8\pi/9}}+\ket{\alpha\,e^{-i2\pi/9}}+\ket{\alpha\,e^{i4\pi/9}}\right].
\end{eqnarray*}
Figure \ref{Wigner3nt0and9} shows the plots of the Wigner function for  the states $\ket{\psi(0)}=\ket{\psi_3}$ and $\ket{\psi(T_{\rm rev}/9)}$. The unitary time evolution operator at $t=T_{\rm rev}/9$ rotates the initial state $45$ degree clockwise direction in the phase space. 

\begin{figure}[h]
\centering
\includegraphics[width=3.0in,height=2.8in]{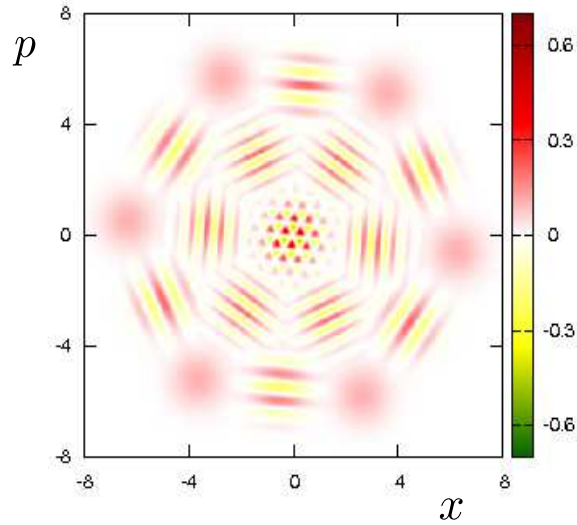} 
\caption{Contour plot of Wigner   function at  $t=T_{\rm rev}/18$ for the  initial  state  $\ket{\psi_3}$ with $\nu=|\alpha|^2=20$. It shows the superposition of two states in the form of $\ket{\psi_3}$.  }
\label{Wigner3Nat18}
\end{figure}

Here, $k$-sub-packet fractional revival occurs at time $t=j\,\trev/9k$ where $j=1, 2, \dots, (9k-1)$ for a given value of   $k (>1)$ with $(j, 9k)=1$.   For example, two sub-packet fractional revival for an initial  state  occurs at $t=T_{\rm rev}/18$ and 
the state at this time is a superposition of two  states of the form $\ket{\psi_3}$:
\begin{eqnarray}
\ket{\psi(\trev/18)}&=&C_1\,N_{3}\left[\ket{\alpha\,e^{-i11\pi/18}}+\ket{\alpha\,e^{i\pi/18}}+\ket{\alpha\,e^{i13\pi/18}}\right]\nonumber\\
&+&C_2\,N_{3}\left[\ket{\alpha\,e^{-i17\pi/18}}+\ket{\alpha\,e^{-i5\pi/8}}+\ket{\alpha\,e^{i7\pi/18}}\right],
\label{psi3at2}
\end{eqnarray}
where $C_1=(1-i)/2$ and $C_2=(1+i)/2$.
Figure \ref{Wigner3Nat18} clearly shows the superposition of two initial states given in equation (\ref{ketpsi3}) with different $\alpha$ values as given in equation (\ref{psi3at2})     at $t=\trev/18$.

\begin{figure}[!htpb]
\centering
\includegraphics[scale=0.45]{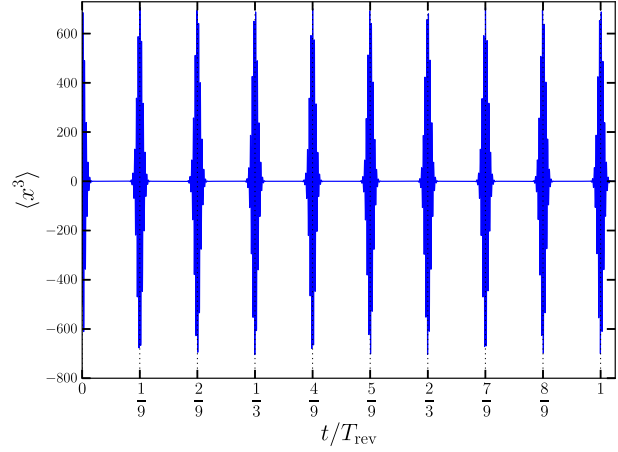} 
\caption{$\aver{x^3(t)}$ as a function of $t/T_{\rm rev}$ for the initial state  $\ket{\psi_3}$ with $\nu=|\alpha|^2=100$. In between $t=0$ and $T_{\rm rev}$,  the third moment of $x$ is constant most of the time except at    $t=jT_{\rm rev}/9$  where  $j=1, 2, \dots, 8$. At these instants  of time the evolved state is a rotated  initial wave packet.}
\label{x3vst3N}
\end{figure}

\begin{figure}[!htpb]
\centering
\includegraphics[scale=0.7]{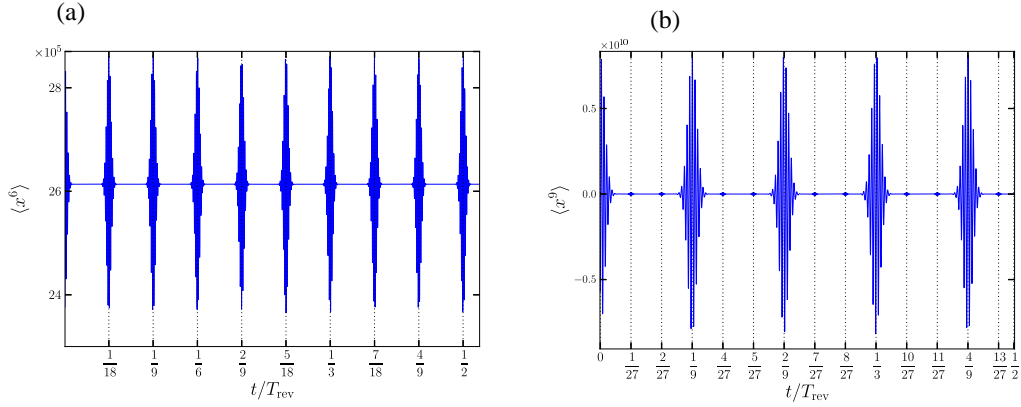} 
\caption{Temporal evolution of higher moments of $x$ for an initial state $\ket{\psi_3}$  with $\nu=|\alpha|^2=100$.   (a) $\aver{x^6(t)}$ is constant most of the time except at   $t=j\,T_{\rm rev}/18$, where $j=1, 2, \dots, 9$. At these instants, $6^{\rm th}$ moment of $x$  shows  a rapid variation, which is a signature of  two-sub-packet fractional revival  and wave packet rotation. (b) $\aver{x^9(t)}$ is constant most of the time except at   $t=j\,T_{\rm rev}/27$, where $j=1, 2, \dots, 13$. At these instants, $9^{\rm th}$ moment of $x$  shows  a rapid variation, which is a signature of  three-sub-packet fractional revival  and wave packet rotation.}
\label{x6x9for3N}
\end{figure}

\begin{figure}[!htpb]
\centering
\includegraphics[width=5.0in,height=3.1in]{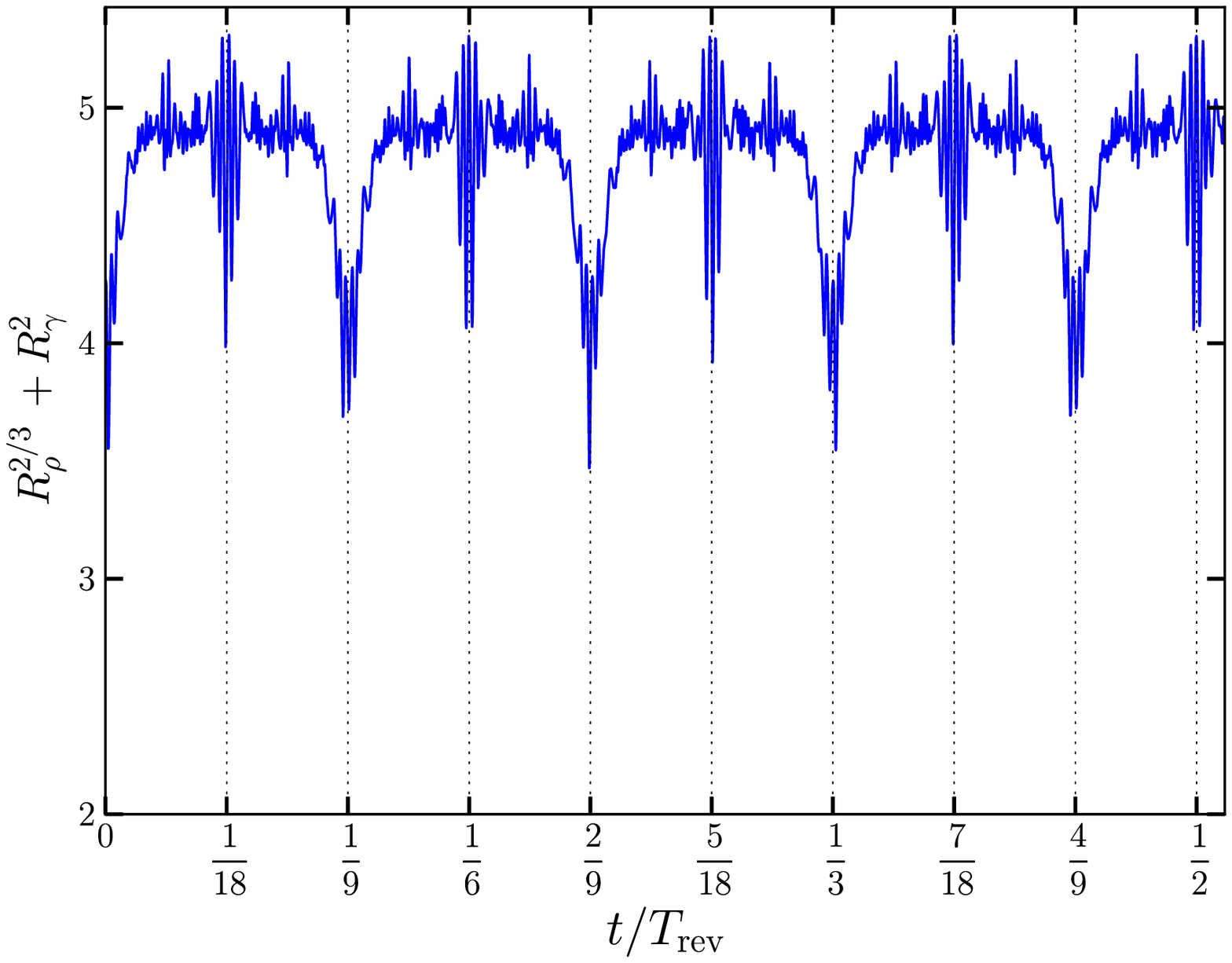} 
\caption{
Time evolution of $R_{\rho}^{(2/3)}+R_{\gamma}^{(2)}$ for the initial state $\ket{\psi_3}$  with $\nu=30$. The main fractional revivals are indicated by vertical dotted lines.}
\label{entropy3N}
\end{figure}

Only the $3k^{\rm th}$ (where $k=1, 2, \dots$) moment of $x$ and $p$ gives non-zero value and all other moments are identically equal to  zero at all times for the initial state $\ket{\psi_3}$. The expectation value of $x^3$  at any time for an initial state $\ket{\psi_3}$ is
\begin{eqnarray}
\aver{x^3(t)}&=&3N_3^{2}\,\nu^{3/2}\, \left[e^{-\nu \left( 1-\cos  6\chi t  \right)}\cos \left( 6\chi t+\nu \sin  6\chi t -{3\pi }/{4} \right)\right.\nonumber\\
&&+\left. e^{-\nu \left( 1-\sin \left( 6\chi t -\pi/6\right) \right)}\cos \left( 6\chi t+\nu \cos \left( 6\chi t+\pi/6 \right)-{3\pi }/{4} \right)\right.\nonumber\\
&&+\left. e^{-\nu \left( 1+\sin \left( 6\chi t +\pi/6\right) \right)}\cos \left( 6\chi t-\nu \cos \left( 6\chi t-\pi/6 \right)-{3\pi }/{4} \right)\right]
\end{eqnarray}
 In between $t=0$ and $\trev$, the above expression for $\aver{x^3}$ is zero most of the times except at  $t=jT_{\rm rev}/9$, where $j=1, 2, \dots, 8$  for  sufficiently large value of $\nu$. These instants correspond to wave packet rotation in phase space.  Figure \ref{x3vst3N} shows the variation of the expectation value $\aver{x^3}$ versus time for the initial state $\ket{\psi_3}$. It shows that   wave packet rotation in phase space is captured in the third moment of $x$.
Expressions for the higher moments of $x$ and $p$ can be deduced readily from the general result
 \begin{eqnarray}
\aver{a^{3k}}&=&3N^2_3\, {\alpha}^{3k}\Big\{e^{-\nu(1-\cos 6k\chi t)} \exp\left[ -3ik(3k-1)\chi t-i\nu \sin 6k\chi t\right] \nonumber\\
&&\left.+ e^{-\nu\left(1-\sin (6k\chi t-\pi/6)\right)} \exp\left[ -3ik(3k-1)\chi t-i\nu \cos (6k\chi t+\pi/6)\right]\right.\nonumber\\
&&+ e^{-\nu\left(1+\sin (6k\chi t+\pi/6)\right)} \exp\left[ -3ik(3k-1)\chi t+i\nu \cos (6k\chi t-\pi/6)\right]\Big\}\nonumber\\
&&
\end{eqnarray}
The time dependence of $3k^{\rm th}$ moments of $x$ is
strongly controlled by the factors $\exp\left[-\nu(1-\cos 6k\chi t)\right]$ and  $\exp\left[-\nu\left(1\pm\sin (6k\chi t-\pi/6)\right)\right], k=1, 2, \dots,$ that modulates the oscillatory term. In between $t=0$ and $t=T_{\rm rev}$, these factors act as a strong damping factor for large values of $\nu$, except at fractional revival times $t=j\trev/9k$. It can be concluded that $k$-sub-packet fractional revivals are captured in the $3k^{\rm th}$ moment of $x$ or $p$. These results are illustrated in figures \ref{x6x9for3N} (a) and \ref{x6x9for3N} (b).
 Figure \ref{x6x9for3N} (a) shows the temporal evolution of the expectation 
value $\aver{x^6(t)}$. We have plotted the graph in between $t=0$ and $\trev/2$ for a better view. It confirms that  sixth moment of $x$  captures the signature of  2-sub-packet fractional revival and rotations.  Figure \ref{x6x9for3N} (b) is a plot of $\aver{x^9(t)}$ versus time which shows the signature of $3$-sub-packet fractional revivals and rotations.  
Figure \ref{entropy3N} shows the  R\'{e}nyi uncertainty  versus time for the initial state $\ket{\psi_3}$ with $\nu=30$. Again, it confirms our analysis based on the expectation values. The main fractional revivals are indicated by vertical dotted lines in the figure. So far we have studied the dynamics of initial states of the
form $\sum_{n}\,C_n \ket{ln}$, where $l=1, 2$ and $3$. In the next section we generalize our results for a general $l$. 

\section{Summary}
\begin{figure}
\centering
\includegraphics[scale=0.45]{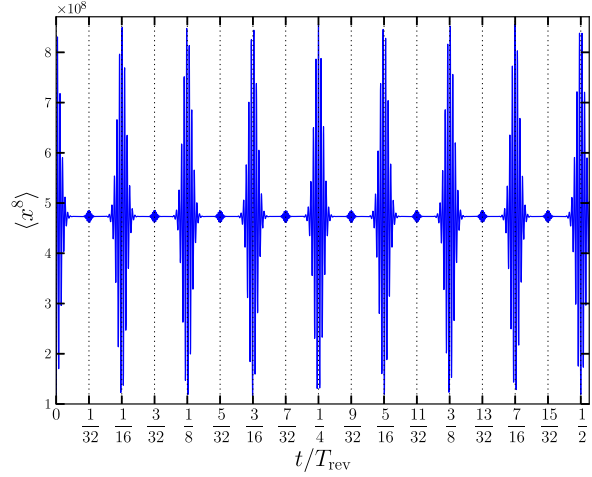} 
\caption{$\aver{x^8(t)}$ as a function of $t/T_{\rm rev}$ for the initial state  $\ket{\psi_4}$ with $\nu=|\alpha|^2=100$.  In between $t=0$ and $t=T_{\rm rev}/2$,  $\aver{x^8(t)}$ is constant most of the time except at   $t=j\,T_{\rm rev}/32$, where $j=1, 2, \dots, 16$. 
At these instants, $8^{\rm th}$ moment of $x$  shows  a rapid variation, which is a signature of  two-sub-packet fractional revival  and wave packet rotation.}
\label{x8for4N}
\end{figure}

\paragraph{}We have extended  the analysis carried out in the above sections  for  an  initial wave packet  $\ket{\psi_l}$ given in equation  (\ref{generaleven}) for a general $l$.  For ready reference, we write it down again  the Fock state representation of $\ket{\psi_l}$:
\begin{eqnarray}
\ket{\psi_{l}}=l\,N_l\,e^{-\nu/2}\,\sum_{n=0}^{\infty} \frac{\alpha^{ln}}{\sqrt{(l\,n)!}}\,\ket{l\,n}.
 \label{generalevennumber2}
 \end{eqnarray}
We investigated the dynamics of initial state  $\ket{\psi_{l}}$ with $l>3$  and found the following general results:
\begin{enumerate}
\item The time evolved state at  $t=j\,\trev/l^2$, where $j=1, 2, \dots, (l^2-1)$,  is a rotated initial wave packet.
\item $k$-sub-packet fractional revival occur at  $t=j\,\trev/l^2k$ where $j=1, 2, \dots, (l^2k-1)$ for a given value of   $k (>1)$ with  $(j,l^2k)=1$.
\item The   distinctive signatures of $k$ sub-packet fractional revivals are captured in $(lk)^{\rm th}$ moments of the operators $x$ and $p$.
\end{enumerate}
We do not write it down the analysis for higher values of $l$ because it is  repetitive, but for completeness we discuss the dynamics of an initial state  $\ket{\psi_4}$. This initial state can be written as a superposition of $4$ coherent states  (see equation (\ref{generaleven})). According to the result (ii) quoted above, two-sub-packet  fractional revival of the initial state $\ket{\psi_4}$ occur at  $t=\,\trev/32$.  Indeed the  initial state $\ket{\psi_4}$  shows two-sub-packet fractional revival at $t=\trev/32$:
\begin{eqnarray*}
 \ket{\psi(\trev/32)}&=N_4 C_1\Big[\ket{\alpha e^{-i 31\pi/32}}+\ket{\alpha e^{-i15\pi/32}}+\ket{\alpha e^{i \pi/32}}+\ket{\alpha e^{i 17\pi/32}}\Big] \nonumber \\
 &+N_4 C_2\Big[\ket{\alpha e^{-i 23\pi/32}}+\ket{\alpha e^{-i7\pi/32}}+\ket{\alpha e^{i 9\pi/32}}+\ket{\alpha e^{i 25\pi/32}}\Big],
 \end{eqnarray*}
where $C_1=(1-i)/2$ and $C_2=(1+i)/2$.
 For the same initial state $\ket{\psi_4}$, Figure \ref{x8for4N}  shows the temporal evolution of the expectation 
value $\aver{x^8(t)}$. We have plotted till $t=\trev/2$ for a better view, i.e., $j$  runs only up to $16$ instead of $31$ in the result (ii).  It  captures the signature of 2-sub-packet fractional revival at $t=j\trev/32$ where $j=1, 2, \dots, 16$ with $(j, 32)=1$  and wave packet rotations at $t=j\trev/16$, where $j=1, 2, \dots, 8$, which confirms our general result (iii) quoted above. 

Experimental manifestations of our results are  possible using continuous-variable quantum-state tomography.   The moments of the operators $x$ and $p$ can be experimentally measured using homodyne correlation techniques with a weak local oscillator \cite{shchukin}.  Wigner function have been reconstructed from the measurements of the quantum statistics of the quadrature operator $x$ \cite{brei,lvov}. Recently, the experimental characterization of photon creation and annihilation operators has been reported \cite{kumar}.  It may possible to measure the R\'{e}nyi entropy using the techniques described in \cite{daley, dmitry}.
\verb``\ack
The authors would like to sincerely  thank the anonymous reviewers   for their valuable comments and suggestions.

\end{document}